\documentclass[11pt,oneside,english,reqno,a4paper]{amsart}

\usepackage{graphicx}
\usepackage{mathrsfs}
\usepackage[numbers]{natbib}
\usepackage[fit]{truncate}
\usepackage{amssymb,amsmath,amsfonts}
\usepackage[english]{babel}
\usepackage[utf8]{inputenc}
\usepackage[compatible]{algpseudocode}
\usepackage{float}
\usepackage{diagbox}
\usepackage[hidelinks]{hyperref}
\usepackage{hhline}
\usepackage{blindtext}
\usepackage{makecell}
\usepackage{caption} 

\newcommand{\R}{\ensuremath\mathbb{R}}

\newcommand{\E}{\ensuremath\mathbb{E}}

\newcommand{\by}{\boldsymbol{y}}
\newcommand{\bbet}{\boldsymbol{\beta}}
\newcommand{\bgam}{\boldsymbol{\gamma}}
\newcommand{\beps}{\boldsymbol{\varepsilon}}
\newcommand{\bx}{\boldsymbol{x}}
\newcommand{\bz}{\boldsymbol{z}}
\newcommand{\bX}{\boldsymbol{X}}
\newcommand{\bZ}{\boldsymbol{Z}}
\newcommand{\bF}{\boldsymbol{F}}
\newcommand{\bone}{\boldsymbol{1}}
\newcommand{\bQ}{\boldsymbol{Q}}
\newcommand{\btet}{\boldsymbol{\vartheta}}
\newcommand{\btau}{\boldsymbol{\tau}}
\newcommand{\bphi}{\boldsymbol{\phi}}
\newcommand{\bet}{\boldsymbol{\eta}}

\newcommand{\blmm}{\texttt{boostLMM} }
\newcommand{\blmma}{\texttt{boostLMM}$^a$ }
\newcommand{\blmmb}{\texttt{boostLMM}$^b$ }

\newenvironment{titlemize}[1]{%
	\paragraph{#1}
	\begin{itemize}}
	{\end{itemize}}

\DeclareMathOperator{\Var}{\text{Var}}
\DeclareMathOperator{\argmax}{arg\,max}
\DeclareMathOperator{\argmin}{arg\,min}

\makeatletter
\def\author@andify{%
	\nxandlist {\unskip ,\penalty-1 \space\ignorespaces}%
	{\unskip {} \@@and~}%
	{\unskip \penalty-2 \space \@@and~}%
}
\makeatother

\numberwithin{equation}{section}

\begin{document}
	
	\title[Addressing cluster-constant covariates]{Addressing cluster-constant covariates in mixed effects models via likelihood-based boosting techniques.}
	\author{Colin Griesbach}
	\author{Andreas Groll}
	\author{Elisabeth Waldmann}
	\thanks{The work on this article was supported by the Interdisciplinary Center for Clinical Research (IZKF) of the Friedrich-Alexander-University Erlangen-Nürnberg (Project J61). Colin Griesbach performed the present work in partial fulfilment of the requirements for obtaining the degree ‘Dr. rer. biol. hum.’ at the Friedrich-Alexander-Universität Erlangen-Nürnberg.\\}
	
	\address{Department of Medical Informatics, Biometry, and Epidemiology, Friedrich-Alexander-Universität Erlangen-Nürnberg, Waldstr. 6, D-91054 Erlangen}
	\email{colin.griesbach@fau.de}
	\address{Fakultät Statistik, Technische Universität Dortmund, D-44221 Dortmund}
	\email{groll@statistik.tu-dortmund.de}
	\address{Department of Medical Informatics, Biometry, and Epidemiology, Friedrich-Alexander-University Erlangen-Nürnberg, Waldstr. 6, D-91054 Erlangen}
	\email{elisabeth.waldmann@fau.de}
	
	\maketitle
	
	\begin{abstract}
		Boosting techniques from the field of statistical learning have grown to be a popular tool for estimating and selecting predictor effects in various regression models and can roughly be separated in two general approaches, namely gradient boosting and likelihood-based boosting. An extensive framework has been proposed in order to fit generalised mixed models based on boosting, however for the case of cluster-constant covariates likelihood-based boosting approaches tend to mischoose variables in the selection step leading to wrong estimates. We propose an improved boosting algorithm for linear mixed models where the random effects are properly weighted, disentangled from the fixed effects updating scheme and corrected for correlations with cluster-constant covariates in order to improve quality of estimates and in addition reduce the computational effort. The method outperforms current state-of-the-art approaches from boosting and maximum likelihood inference which is shown via simulations and various data examples.
	\end{abstract}
	
	\section*{Introduction}
	Linear mixed models \cite{Laird.1982} proved to be a very popular tool for analysing data with repeated measurements, especially clustered longitudinal data from clinical surveys. In order to use mixed models for prediction analysis various approaches to regularised regression like lasso \cite{Friedman.2010, Tibshirani.1996} and boosting techniques \cite{Freund.1996} have been proposed. Boosting in general can be distinguished between gradient boosting \cite{Breiman.1998, Breiman.1999} and likelihood-based boosting \cite{Tutz.2006, Tutz.2007}. Both boosting methods are capable of fitting mixed models and for the latter an extensive framework has been proposed towards this matter in \cite{Tutz.2010, Groll.2012, Tutz.2013} and is included in the R package \texttt{GMMBoost} \cite{Groll.2013} available on \texttt{CRAN}. Apart from improving prediction analysis, boosting methods are due to an iterative and component-wise fitting process suitable for high dimensional data and implicitly offer variable selection.
	
\begin{figure}
		\centering
		\includegraphics[width = 12cm, trim = 1cm 0 0 0]{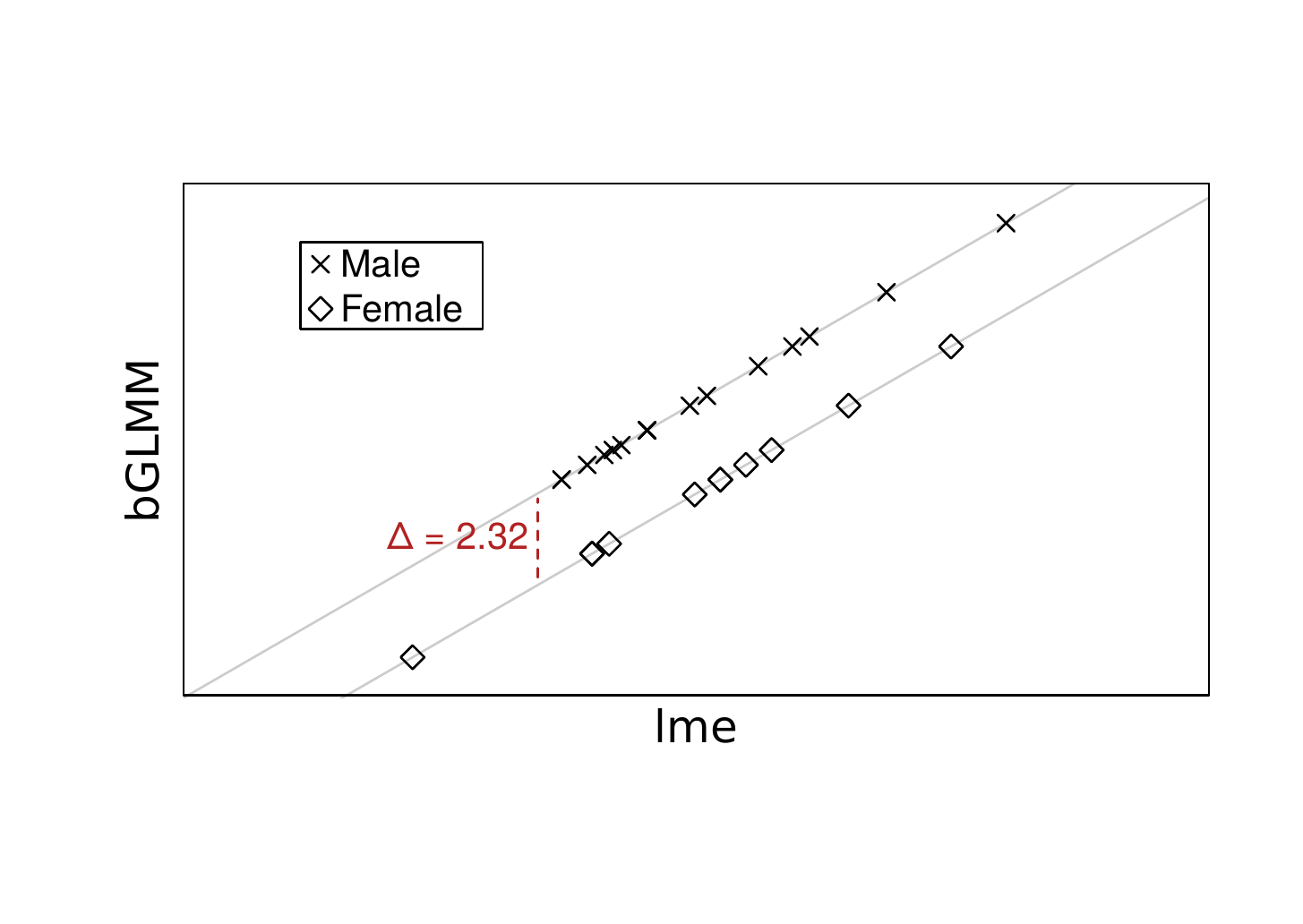}
		\caption{Comparison between random intercept estimates by \texttt{lme} and \texttt{bGLMM}.}
		\label{fig_problem}
	\end{figure}

	However, the \texttt{bGLMM} algorithm from the \texttt{GMMBoost} package tends to struggle with cluster-constant covariates, e.g.\ baseline covariates like gender or treatment group in longitudinal studies. The specified selection and updating procedure of the \texttt{bGLMM} algorithm tends to favour cluster-varying covariates while the simultaneously updated random intercepts partly account for effects actually evolving from cluster-constant covariates. As shown in Figure \ref{fig_problem}, this malfunction already occurs in a very basic data example with the popular Orthodont dataset, which is, among others, available in the \texttt{nlme} package. The dataset depicts the evolution of an orthodontal measurement of 27 children and contains two covariates. A basic linear mixed model with random intercepts returns the two coefficient estimates $\hat{\beta}_\text{gender}^\text{lme} = -2.32$ by \texttt{lme} and $\hat{\beta}_\text{gender}^\text{b} = 0.00$ by \texttt{bGLMM} for the effect of the cluster-constant covariate gender. The reason for this difference becomes clear when looking at the random intercepts, where \texttt{bGLMM} tends to compensate the missing effect for gender by assigning every female subject a random intercept lowered by $2.32$.
	
	We propose an updated algorithm with various changes in order to avoid the phenomenon of random intercepts growing too quickly. These changes include the usage of smaller starting values and weaker random-effects updates to prevent the random effects from growing too fast as well as undocking the random effects update from the fixed effects boosting scheme, which guarantees a fair comparison between the single covariates for the fixed effects. Most importantly, we introduce a correction step for the random effects estimation to avoid possible correlations with observed covariates.
	
	The remainder of the paper is structured as follows: Section \ref{sec_methods} formulates the underlying model and the updated boosting algorithm as well as a detailed discussion of the changes. The algorithm is then evaluated and compared using a simulation study described in Section \ref{sec_simulation} and applied to real world data examples in Section \ref{sec_data}. Finally, the results and possible extensions are discussed.
	
	\section{Methods}\label{sec_methods}
	
	\subsection{Model Specification}\label{ssec_model}
	For clusters $i = 1,\dots,n$ with observations $j = 1,\dots,n_i$ we consider the linear mixed model
	\begin{equation*}
	y_{ij} = \beta_0 + \bx_{ij}^T \bbet + \bz_{ij}^T \bgam_i + \varepsilon_{ij},
	\end{equation*}
	with covariate vectors $\bx_{ij}^T = (x_{ij1},\dots,x_{ijp})$ and $\bz_{ij}^T = (z_{ij1},\dots,z_{ijq})$ referring to the fixed and random effects $\bbet$ and $\bgam_i$, respectively. The random components are assumed to follow normal distributions, i.e. $\varepsilon_{ij} \sim \mathcal{N}(0, \sigma^2)$ for the model error and $\bgam_i \sim \mathcal{N}^{\otimes q}(\boldsymbol{0}, \bQ)$ for the random effects. This leads to a cluster-wise notation
	\begin{equation*}
	\by_i = \beta_0 \bone + \bX_i \bbet + \bZ_i \bgam_i + \beps_i
	\end{equation*}
	with $\by_i = (y_{i1}, \dots, y_{in_i})^T$, $\bone = (1,\dots,1)$, $\bX_i = (\bx_{i1},\dots,\bx_{in_i})^T$, $\bZ_i = (\bz_{i1},\dots,\bz_{in_i})^T$ and $\beps_i = (\varepsilon_{i1},\dots,\varepsilon_{in_i})$. Finally, we get the common matrix notation
	\begin{equation}\label{eq_full_model}
	\by = \beta_0 \bone + \bX \bbet + \bZ \bgam + \beps
	\end{equation}
	of the full model with observations $\by = (\by_1^T,\dots,\by_n^T)^T$, design matrices $\bX = [\bX_1^T,\dots,\bX_n^T]^T$ and the block-diagonal $\bZ = diag(\bZ_1, \dots,\bZ_n)$. The random components $\beps = (\beps_1^T,\dots,\beps_n^T)^T$ and $\bgam = (\bgam_1^T,\dots,\bgam_n^T)^T$ have corresponding covariance matrices $\sigma^2 \boldsymbol{I}_N$ and $diag(\bQ,\dots,\bQ)$ where $\boldsymbol{I}_N$ is the $N = \sum n_i$ dimensional unit matrix.
	
	In order to perform likelihood inference, let $\btet = (\beta_0, \bbet^T, \bgam^T)$ denote the effects and $\bphi = (\sigma^2, \btau)$ information of the random components, where $\btau$ contains the values of $\bQ$. The log-likelihood of the model is
	\begin{equation*}
	\ell(\btet, \bphi) = \sum_{i = 1}^{n} \log \int f(\by_i|\btet, \bphi) p(\bgam_i|\bphi) d \bgam_i,
	\end{equation*}
	where $f(\cdot|\btet, \bphi)$ and $p(\cdot|\bphi)$ denote the normal densities of the model error and the random effects. Laplace approximation following \cite{Breslow.1993} results  in the penalized log-likelihood
	\begin{equation}\label{eq_likelihood}
	\ell^\text{pen} (\btet, \bphi) = \sum_{i = 1}^{n} \log f(\by_i|\btet, \bphi) - \frac12 \sum_{i = 1}^{n} \bgam_i^T \bQ^{-1} \bgam_i,
	\end{equation}
	which is going to be maximized simultaneously for $\btet$ and $\bphi$ by likelihood-based boosting-techniques discussed in the following subsection.
	
	\subsection{Boosting Algorithm}
	The following algorithm maximizes the likelihood (\ref{eq_likelihood}) corresponding to the linear mixed model (\ref{eq_full_model}) via component-wise likelihood-based boosting.
	In order to correct  the random effects for cluster-constant covariates, let $\bX_\text{c} \in \text{Mat}_\R(n, p_\text{c})$ denote the design matrix of the $p_\text{c} \leq p$ covariates, which stay constant within a cluster and $\bgam_{\bullet s} \in \R^n$ the vector of the $s$th random effects for all $n$ clusters. Furthermore set $\tilde{\bX}_\text{c} = (\bone, \bX_\text{c})$ and compute the correction matrix $\bX_\text{cor} = (\tilde{\bX}_\text{c}^T \tilde{\bX}_\text{c})^{-1} \tilde{\bX}_\text{c}^T$.
	\ \\
	
	\noindent\rule[0.5ex]{\linewidth}{1pt}
	
	\begin{titlemize}{\textbf{Algorithm} \blmm}
		\item \textbf{Initialize} estimates with starting values $\hat{\btet}^{[0]}$ and $\hat{\bphi}^{[0]}$. Choose total number of iterations $m_\text{stop}$ and step length $\nu$.
		\item \textbf{for} $m=1$ to $m_\text{stop}$ \textbf{do}
		
		\item[] \textbf{step1: Update fixed effects}\\
		For $r = 1, \dots, p$ define $\bbet_r := (\hat{\beta}_0^{[m-1]}, \hat{\beta}_r^{[m-1]})^T$ with $\hat{\beta}_r^{[m-1]}$ denoting the $r$th component of $\hat{\bbet}^{[m-1]}$. Compute score vector and Fisher matrix
		\begin{equation*}
		\boldsymbol{s}_r(\bbet_r) = \frac{\partial \ell^\text{pen}}{\partial \bbet_r}, \quad \boldsymbol{F}_r(\bbet_r) = -\E\left[\frac{\partial^2 \ell^\text{pen}}{\partial \bbet_r\partial \bbet_r^T}\right]
		\end{equation*}
		with respect to the current intercept $\hat{\beta}_0^{[m-1]}$ and the $r$th linear effect $\hat{\beta}_r^{[m-1]}$. Obtain $p$ possible updates
		\begin{equation*}
		\boldsymbol{u}_r = \boldsymbol{F}_r(\bbet_r)^{-1} \boldsymbol{s}_r(\bbet_r) \in \R^2
		\end{equation*}
		and find the best performing component $* \in \{1,\dots,p\}$ maximizing the unpenalized likelihood. This yields the update $\boldsymbol{u}_* = (u_0, u_*)$ containing the update $u_*$ for the effect $*$ with corresponding intercept update $u_0$. Receive $\hat{\beta}_0^{[m]}$, $\hat{\bbet}^{[m]}$ by updating
		\begin{align}\label{eq_weak_update}
		\begin{split}
		\hat{\beta}_0^{[m]} &= \hat{\beta}_0^{[m-1]} + \nu u_0,\\
		\hat{\beta}_r^{[m]} &= \begin{cases}
		\hat{\beta}_r^{[m-1]} &\text{if } r \neq *,\\
		\hat{\beta}_r^{[m-1]} + \nu u_* &\text{if } r = *,
		\end{cases}
		\quad r = 1,\dots,p.
		\end{split}
		\end{align}
		
		\item[] \textbf{step2: Update random effects}\\
		Receive a first update
		\begin{equation*}
		\hat{\bgam}^{[m-1]} \rightarrow \tilde{\bgam}^{[m]}
		\end{equation*}
		for random effects in an additional Fisher scoring step based on the penalized log-likelihood $\ell^\text{pen}$. Then correct this update for cluster-constant covariates by
		\begin{equation*}
		\hat{\bgam}_{\bullet s}^{[m]} = \begin{cases}
		\tilde{\bgam}_{\bullet s}^{[m]} - \bX_\text{cor} \tilde{\bgam}_{\bullet s}^{[m]}, &\text{if } s = 1,\\
		\tilde{\bgam}_{\bullet s}^{[m]} - m(\tilde{\bgam}_{\bullet s}^{[m]}), &\text{if } s = 2,\dots,q.
		\end{cases}
		\end{equation*}
		We set without loss of generality $s=1$ as the random intercept component and $m(\cdot)$ as the arithmetic mean.
		
		\item[] \textbf{step3: Update variance-covariance-components}\\
		Update variance-covariance-components
		\begin{equation*}
		\hat{\sigma}^{2[m-1]} \rightarrow \hat{\sigma}^{2[m]}, \quad \hat{\bQ}^{[m-1]} \rightarrow \hat{\bQ}^{[m]}
		\end{equation*}
		using an approximate EM-algorithm.
		\item[] \textbf{end for}
		
		\item \textbf{Stop} the algorithm at the best performing $m_*$ with respect to quality of prediction. Return $\hat{\btet}^{[m_*]}$ and $\hat{\bphi}^{[m_*]}$ as the final estimates.
	\end{titlemize}
	
	\noindent\rule[0.5ex]{\linewidth}{1pt}
	
	\subsection{Computational details of the Algorithm}
	We give a stepwise description of the computational details of the \blmm algorithm. For simplicity, we omit iteration indices and hats indicating estimated values whenever appropriate.
	
	\textbf{Starting values.} The parameters actually underlying the boosting process are necessarily set to zero, thus $\hat{\bbet}^{[0]} = \boldsymbol{0}$. There exist two natural options for intercept and random effects together with variance-covariance-components. The first one is fitting a standard linear mixed model for intercept and random effects
	\begin{equation}\label{eq_start_int_model}
	\by = \beta_0 \bone + \bZ \bgam + \beps
	\end{equation}
	by using e.g. the function \texttt{lme} from the R package \texttt{nlme} and extracting the starting values from the model fit. The second one is setting $\hat{\beta}_0^{[0]} = m(\by)$, $\hat{\sigma}^{2[0]} = \Var(\by)$ and initializing random effects $\hat{\bgam}^{[0]} = \boldsymbol{0}$ to be zero with small covariance-matrix, e.g.\ $\hat{\boldsymbol{Q}}^{[0]} = diag(0.1,\dots, 0.1)$.
	
	\textbf{Fixed effects boosting process.} The computation of the $r$th update is straight forward by calculating
	\begin{equation*}
	\boldsymbol{s}_r(\bbet_r) = \sigma^{-2}\tilde{\bX}_r^T (\by - \bet), \quad \boldsymbol{F}_r(\bbet_r) =  \sigma^{-2}\tilde{\bX}_r^T \tilde{\bX}_r,
	\end{equation*}
	where $\tilde{\bX} = (\bone, \bX_{\bullet r})$ is a $N\times 2$ matrix containing a column of ones and the $r$th column of $\bX$ associated with the $r$th covariate and $\bet$ denoting the current fit. This leads to $p$ possible parameter vectors $\hat{\btet}_r$, where only the intercept and $r$th component received a full, i.e.\ not scaled by $\nu$, update according to $\boldsymbol{u}_r$. The best performing component is then determined by finding
	\begin{equation*}
	* = \underset{r = 1,\dots,p}\argmax \ \sum_{i = 1}^{n} \log f(\by_i|\hat{\btet}_r, \hat{\bphi})
	\end{equation*}
	and receives an actual weak update as depicted in updating scheme (\ref{eq_weak_update}).
	
	\textbf{Random effects update.} This is the first of two steps for receiving updated estimates for the random effects. In the beginning, an uncorrected update for the random effects $\bgam$ is obtained by calculating
	\begin{equation*}
	\boldsymbol{s}_\text{ran}(\bgam) = \frac{\partial \ell^\text{pen}}{\partial \bgam}, \quad \boldsymbol{F}_\text{ran}(\bgam) = -\E\left[\frac{\partial^2 \ell^\text{pen}}{\partial \bgam\partial \bgam^T}\right]
	\end{equation*}
	and weakly updating
	\begin{equation*}
	\tilde{\bgam}^{[m]} = \hat{\bgam}^{[m-1]} + \nu\boldsymbol{F}_\text{ran}(\bgam)^{-1}\boldsymbol{s}_\text{ran}(\bgam).
	\end{equation*}
	Note that this differs from the approach in \cite{Tutz.2010} as the random effects are updated separately and in addition also receive an update scaled by the step length $\nu$. The weak update ensures that the random effects don't grow to quickly compared to the fixed effects. The disentanglement of the random effects update from the fixed effects updating scheme on the other hand guarantees a fair comparison of the single fixed effects, where the random effects do not play a crucial role. In addition the Fisher matrix
	\begin{equation*}
	\boldsymbol{F}_\text{ran}(\bgam) = diag(\bF_1,\dots,\bF_n), \quad \bF_i = \sigma^{-2}\bZ_i^T\bZ_i + \bQ^{-1}
	\end{equation*}
	has block-diagonal form making the inversion much easier and thus strongly reducing the computational effort.
	
	\textbf{Random effects correction.} Although the model is uniquely identifiable with the added penalty for the random effects, a naive boosting approach tends to converge into local optima of the penalized log-likelihood and the problem of too strongly growing random effects, which has been mentioned in the introduction, occurs. While weakened and disentangled updates for the random effects improve this issue, an additional correction is needed in order to prevent it completely. Hence, instead of using the unaltered random intercept estimate $\tilde{\bgam}_{\bullet 1}^{[m]}$ we proceed with the orthogonalised estimates
	\begin{equation*}
	\hat{\bgam}_{\bullet 1}^{[m]} = \tilde{\bgam}_{\bullet 1}^{[m]} - (\tilde{\bX}_\text{c}^T \tilde{\bX}_\text{c})^{-1} \tilde{\bX}_\text{c}^T\tilde{\bgam}_{\bullet 1}^{[m]},
	\end{equation*}
	which result by counting out the orthogonal projections of $\tilde{\bgam}_{\bullet 1}^{[m]}$ onto the subspace generated by the cluster-constant covariates $\tilde{\bX}_\text{c}$. This ensures that the resulting estimates $\hat{\bgam}_{\bullet 1}^{[m]}$ are uncorrelated with any cluster-constant covariates.
	
	\textbf{Updating variance-covariance-components.} The covariance matrix $\bQ$ of the random effects is updated with an approximate EM-algorithm using the posterior curvatures $\boldsymbol{F}_{i}$ of the random effects model \cite{Fahrmeir.2001}. An update is received by computing
	\begin{equation*}
	\hat{\bQ} = \frac1n \sum_{i=1}^n \left(\boldsymbol{F}_{i}^{-1} + \hat{\bgam_i}\hat{\bgam}_i^T\right).
	\end{equation*}
	The current longitudinal model error is obtained by finding
	\begin{equation*}
	\hat{\sigma}^2 = \underset{\sigma^2 > 0}\argmax \ \sum_{i = 1}^{n} \log f(\by_i|\btet, (\sigma^2, \hat{\btau}))
	\end{equation*}
	using the R Base function \texttt{optimize}.
	
	\textbf{Stopping iteration.} For $m \to \infty$ the algorithm would eventually converge to the regular maximum likelihood estimate. The procedure is stopped early according to quality of prediction which implicitly offers variable selection. While likelihood-based boosting algorithms rely on information criteria like AIC and BIC \cite{Akaike.1973, Schwarz.1978}, we use k-fold cross validation following \cite{Mueller.2013}. Set $\bQ^* = \sigma^{-2}\bQ$. The data is cluster-wise partitioned into $k$ fairly equal subsets to compute
	\begin{equation*}
	CV_k^{[m]} = \frac1k\sum_{l = 1}^k \frac1{N_l}\left(\by_l - \bX_l \hat{\bbet}_{-l}^{[m]}\right)^T \left(\boldsymbol{I}_{N_l} + \bZ_l \hat{\bQ}_{-l}^{*[m]}\bZ_l\right)^{-1}\left(\by_l - \bX_l \hat{\bbet}_{-l}^{[m]}\right)
	\end{equation*}
	for every iteration $m = 1,\dots,m_\text{stop}$, where $N_l$ observations $\by_l$, $\bX_l$ and $\bZ_l$ of one subset $l = 1,\dots,k$ are used to evaluate the estimates $\hat{\bbet}_{-l}^{[m]}$, $\hat{\sigma}_{-l}^{2[m]}$ and $\hat{\bQ}_{-l}^{[m]}$ after $m$ iterations based on the remaining data. For $k=n$ this is asymptotically equivalent to the marginal AIC, which has been proved in \cite{Fang.2011}. Averaged over all $k$ folds we then obtain $m_*$ by
	\begin{equation*}
	m_* = \underset{m = 1,\dots,m_\text{stop}} \argmin CV_k^{[m]}.
	\end{equation*}
	
	\section{Simulations}\label{sec_simulation}
	The algorithm is evaluated with a simulation study. The primary focus is to show that the algorithm solves the identification problem of the random effects and thus is compared to the \texttt{bGLMM} function of the \texttt{GMMBoost} package available on \texttt{CRAN}. Furthermore, its performance is compared to the classical method implemented in the \texttt{lme} function of the \texttt{nlme} package with respect to accuracy of estimates, variable selection properties and high dimensionality. As a side note we also report the elapsed computation time.
	
	\subsection{Setup}
	For $i = 1,\dots,50$ and $j = 1,\dots,10$ we consider the setup
	\begin{equation*}
	y_{ij} = \beta_0 + \beta_1 x_{i1} + \beta_2 x_{i2} + \beta_3 x_{ij3} + \beta_4 x_{ij4} + \sum_{r=5}^{p} \beta_rx_{ijr} + \gamma_{0i} + \varepsilon_{ij}
	\end{equation*}
	with values $\beta_0 = 1$, $\beta_1 = 2$, $\beta_2 = 4$, $\beta_3 = 3$ and $\beta_4 = 5$ for the fixed effects, $x_{ir}, x_{ijr} \sim \mathcal{N}(0,1)$ for the cluster-constant and cluster-varying covariates and $\gamma_{0i} ~ \sim \mathcal{N}(0, \tau^2)$ and $\varepsilon_{ij} \sim \mathcal{N}(0, \sigma^2)$ for the random components with $\sigma = 0.4$ and $\tau \in \{0.4, 0.8, 1.6\}$. The total amount of covariates is evaluated for the three different cases $p \in \{10, 50, 500\}$ ranging from low to high dimensional setups.
	
	For $\bbet = (\beta_0,\dots,\beta_p)^T$ we consider mean squared errors
	\begin{equation*}
	\text{mse}_{\bbet} := \|\bbet - \hat{\bbet}\|^2, \quad \text{mse}_\tau := (\tau-\hat{\tau})^2
	\end{equation*}
	as an indicator for estimation accuracy. Variable selection properties are evaluated by calculating the false positives rates, i.e.\ the rate of noninformative covariates being selected. Finally, the elapsed time is measured in seconds where each simulation run was carried out on a \textit{2 x 2.66 GHz-6-Core Intel Xeon} CPU without any parallelisation. In total we compare four different routines. Additional to \texttt{lme} and \texttt{bGLMM} we use two versions of \blmm with varying starting values. Here \blmma denotes the algorithm with initial random effects set to zero and \blmmb the version with random effects estimates as starting values as described in formula (\ref{eq_start_int_model}). Note that the gradient-boosting alternative \texttt{mboost} \cite{Hothorn.mboost} was not included since it is not capable of estimating the variance-covariance-components of the random effects.

	\subsection{Results}
\begin{figure}[h]
		\centering
		\includegraphics[width = 12cm, trim = 1cm 0 0 0]{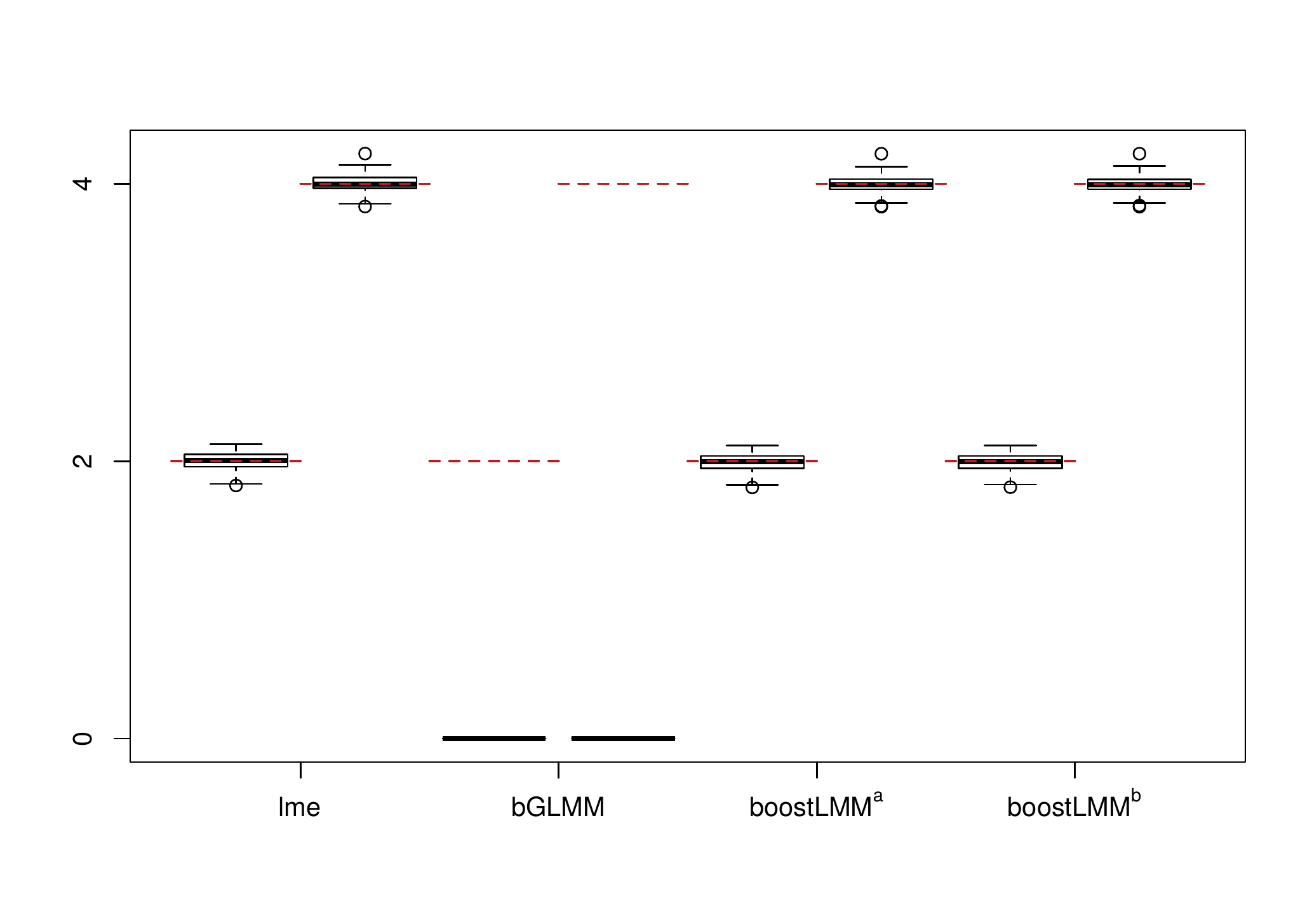}
		\caption{Effect estimates for $\hat{\beta}_1$ (left) and $\hat{\beta}_2$ (right) with $p=10$, $\tau = 0.4$.}
		\label{fig_estimates}
	\end{figure}

	Table \ref{tab_sim_rint_mseb} depicts the results for $\text{mse}_{\bbet}$ as well as variable selection properties. In general, \blmma and \blmmb show no noticeable differences and outperform \texttt{lme} and \texttt{bGLMM} with respect to estimation accuracy, especially when dimensions increase. Note that the high error rates for \texttt{bGLMM} result from the corrupt random intercepts which prevent the coefficients $\hat{\beta}_1$ and $\hat{\beta}_2$ from being selected almost every time and thus lead to an error of $2^2+4^2=20$ which can be also seen in Figure \ref{fig_estimates}.
	Variable selection performs decently, although the information criterion based selection process of \texttt{bGLMM} clearly shows its advantages. This of course comes with the price of very high computational effort as shown in Table \ref{tab_sim_rint_mt}. Note that in contrast to \texttt{lme} the \texttt{bGLMM} algorithm is technically capable of estimating very high dimensional datasets where the number of covariates exceeds the number of total measurements. In practise however the tremendous computational effort did not allow to execute the code in a reasonable time span, which is why there are no results for both \texttt{lme} as well as \texttt{bGLMM} in the $p=500$ setup.
	
	Estimates for the variance-covariance-structure are shown in Table \ref{tab_sim_rint_mset}. Again, the high error rates for \texttt{bGLMM} are due to the random intercepts which obviously result in a random effects variance being too large. Noticeably, both \blmm versions tend to have a slightly better accuracy for the random components than \texttt{lme}.
	
	\begin{table}
		\centering
		\begin{tabular}{ccccccccccccc}
			\hline
			&&&\texttt{lme}&&\multicolumn{2}{c}{\texttt{bGLMM}}&&\multicolumn{2}{c}{\blmma}&&\multicolumn{2}{c}{\blmmb}\\
			$\tau$&p&&$\text{mse}_{\bbet}$&&$\text{mse}_{\bbet}$&f.p.&&$\text{mse}_{\bbet}$&f.p.&&$\text{mse}_{\bbet}$&f.p.\\
			\hline
			0.4 & 10 && 0.014 && 20.463 & 0.01 && 0.013 & 0.62 && 0.013 & 0.62 \\
			0.4 & 50 && 0.031 && 20.463 & 0.01 && 0.018 & 0.46 && 0.18 & 0.45 \\
			0.4 & 500 && \textbf{-} && \textbf{-} & \textbf{-} && 0.027 & 0.16 && 0.027 & 0.16 \\
			\hline\hline
			0.8 & 10 && 0.046 && 20.437 & 0.01 && 0.045 & 0.53 && 0.045 & 0.52 \\
			0.8 & 50 && 0.062 && 20.437 & 0.01 && 0.048 & 0.30 && 0.048 & 0.30 \\
			0.8 & 500 && \textbf{-} && \textbf{-} & \textbf{-} && 0.056 & 0.11 && 0.056 & 0.11 \\
			\hline\hline
			1.6 & 10 && 0.174 && 20.476 & 0.01 && 0.174 & 0.46 && 0.174 & 0.48 \\
			1.6 & 50 && 0.191 && 20.476 & 0.01 && 0.176 & 0.24 && 0.176 & 0.24 \\
			1.6 & 500 && \textbf{-} && \textbf{-} & \textbf{-} && 0.183 & 0.07 && 0.182 & 0.07 \\
		\end{tabular}
		\caption{Results for $\text{mse}_{\bbet}$ and false positives.}
		\label{tab_sim_rint_mseb}
	\end{table}
	
	\begin{table}
		\centering
		\begin{tabular}{cccccccccc}
			\hline
			&&&\texttt{lme}&&\texttt{bGLMM}&&\blmma&&\blmmb\\
			$\tau$&p&&$\text{mse}_\tau$&&$\text{mse}_\tau$&&$\text{mse}_\tau$&&$\text{mse}_\tau$\\
			\hline
			0.4 & 10 && 0.001 && 19.331 && 0.001 && 0.001 \\
			0.4 & 50 && 0.001 && 19.331 && 0.001 && 0.001 \\
			0.4 & 500 && \textbf{-} && \textbf{-} && 0.001 && 0.001 \\
			\hline\hline
			0.8 & 10 && 0.020 && 15.756 && 0.019 && 0.019 \\
			0.8 & 50 && 0.020 && 15.756 && 0.018 && 0.018 \\
			0.8 & 500 && \textbf{-} && \textbf{-} && 0.019 && 0.019 \\
			\hline\hline
			1.6 & 10 && 0.306 && 5.207 && 0.289 && 0.289 \\
			1.6 & 50 && 0.307 && 5.207 && 0.288 && 0.288 \\
			1.6 & 500 && \textbf{-} && \textbf{-} && 0.289 && 0.289 \\
		\end{tabular}
		\caption{Results for $\text{mse}_\tau$.}
		\label{tab_sim_rint_mset}
	\end{table}
	
	\begin{table}
		\centering
		\begin{tabular}{ccccccccccccc}
			\hline
			&&&\texttt{lme}&&\multicolumn{2}{c}{\texttt{bGLMM}}&&\multicolumn{2}{c}{\blmma}&&\multicolumn{2}{c}{\blmmb}\\
			$\tau$&p&&time&&$m_*$&time&&$m_*$&time&&$m_*$&time\\
			\hline
			0.4 & 10 && 0.08 && 144 & 879 && 310 & 120 && 313 & 121 \\
			0.4 & 50 && 0.26 && 144 & 4146 && 373 & 260 && 370 & 263 \\
			0.4 & 500 && \textbf{-} && \textbf{-} & \textbf{-} && 486 & 8955 && 487 & 9079 \\
			\hline\hline
			0.8 & 10 && 0.08 && 144 & 881 && 273 & 119 && 278 & 120 \\
			0.8 & 50 && 0.23 && 144 & 4149 && 284 & 256 && 288 & 258 \\
			0.8 & 500 && \textbf{-} && \textbf{-} & \textbf{-} && 375 & 8693 && 372 & 8746 \\
			\hline\hline
			1.6 & 10 && 0.08 && 144 & 866 && 260 & 119 && 267 & 120 \\
			1.6 & 50 && 0.24 && 144 & 4126 && 261 & 255 && 264 & 258 \\
			1.6 & 500 && \textbf{-} && \textbf{-} & \textbf{-} && 297 & 8561 && 297 & 8618 \\
		\end{tabular}
		\caption{Results for optimal stopping iteration $m_*$ and elapsed computation time in seconds.}
		\label{tab_sim_rint_mt}
	\end{table}
	
	\subsection{Random Slopes}
	We now consider a slightly altered setup with added random slopes for the two informative cluster-varying covariates, i.e.
	\begin{align*}
	y_{ij} &= \beta_0 + \beta_1 x_{i1} + \beta_2 x_{i2} + \beta_3 x_{ij3} + \beta_4 x_{ij4} + \sum_{r=5}^{p} \beta_rx_{ijr}\\
	&\quad + \gamma_{0i} + \gamma_{1i}x_{ij3} + \gamma_{2i}x_{ij4} + \varepsilon_{ij}
	\end{align*}
	with
	\begin{equation*}
	(\gamma_{0i}, \gamma_{1i}, \gamma_{2i}) \sim \mathcal{N}^{\otimes 3}(\boldsymbol{0}, \bQ), \quad \bQ := \begin{pmatrix}
	\tau^2 & \tau^* & \tau^*\\
	\tau^* & \tau^2 & \tau^*\\
	\tau^* & \tau^* & \tau^2
	\end{pmatrix},
	\end{equation*}
	where $\tau \in \{0.4, 0.8, 1.6\}$ and $\tau^*$ is chosen so that $cor(\gamma_{ki}, \gamma_{li}) = 0.6$ for all $k,l = 1,2,3$ holds. We evaluate the mean squared errors
	\begin{equation*}
	\text{mse}_{\bbet} := \|\bbet - \hat{\bbet}\|^2, \quad \text{mse}_{\bQ} := \|\bQ-\hat{\bQ}\|_F^2
	\end{equation*}
	with $\|\cdot\|_F$ denoting the Frobenius norm of a given matrix.
	
	Tables \ref{tab_sim_rslp_mseb} and \ref{tab_sim_rslp_mseQ} contain the results for mean squared errors and false positive rates in the case of random slopes. In most setups the versions of \blmm have a slightly better error rate than \texttt{lme} while again the different starting values do not seem to have an impact on the results. Similar to the random intercept model, variable selection properties are improving with increased amount of dimensions, but since the algorithm with its optimal stopping iteration being determined by cross validation is not specifically trained for variable selection, the false positives rates are not as good as for procedures relying on information criteria like \texttt{bGLMM}. Similar to the fixed effects, \blmm outperforms \texttt{lme} with respect to estimation accuracy of the covariance structure with lower error rates in every single setup.
	
	\begin{table}
		\centering
		\begin{tabular}{ccccccccccccc}
			\hline
			&&&\texttt{lme}&&\multicolumn{2}{c}{\texttt{bGLMM}}&&\multicolumn{2}{c}{\blmma}&&\multicolumn{2}{c}{\blmmb}\\
			$\tau$&p&&$\text{mse}_{\bbet}$&&$\text{mse}_{\bbet}$&f.p.&&$\text{mse}_{\bbet}$&f.p.&&$\text{mse}_{\bbet}$&f.p.\\
			\hline
			0.4 & 10 && 0.018 && 29.845 & 0.00 && 0.020 & 0.61 && 0.020 & 0.62 \\
			0.4 & 50 && 0.039 && 29.847 & 0.01 && 0.026 & 0.49 && 0.026 & 0.49 \\
			0.4 & 500 && \textbf{-} && \textbf{-} & \textbf{-} && 0.034 & 0.14 && 0.034 & 0.14 \\
			\hline\hline
			0.8 & 10 && 0.059 && 29.986 & 0.00 && 0.069 & 0.43 && 0.069 & 0.47 \\
			0.8 & 50 && 0.081 && 29.990 & 0.01 && 0.073 & 0.34 && 0.073 & 0.35 \\
			0.8 & 500 && \textbf{-} && \textbf{-} & \textbf{-} && 0.083 & 0.13 && 0.083 & 0.13 \\
			\hline\hline
			1.6 & 10 && 0.217 && 30.380 & 0.00 && 0.265 & 0.38 && 0.265 & 0.43 \\
			1.6 & 50 && 0.239 && 30.387 & 0.01 && 0.269 & 0.25 && 0.268 & 0.26 \\
			1.6 & 500 && \textbf{-} && \textbf{-} & \textbf{-} && 0.277 & 0.11 && 0.278 & 0.11 \\
		\end{tabular}
		\caption{Results for $\text{mse}_{\bbet}$ and false positives with random slopes.}
		\label{tab_sim_rslp_mseb}
	\end{table}
	
	\begin{table}
		\centering
		\begin{tabular}{cccccccccc}
			\hline
			&&&\texttt{lme}&&\texttt{bGLMM}&&\blmma&&\blmmb\\
			$\tau$&p&&$\text{mse}_{\bQ}$&&$\text{mse}_{\bQ}$&&$\text{mse}_{\bQ}$&&$\text{mse}_{\bQ}$\\
			\hline
			0.4 & 10 && 0.009 && 556.455 && 0.009 && 0.009 \\
			0.4 & 50 && 0.009 && 556.469 && 0.009 && 0.009 \\
			0.4 & 500 && \textbf{-} && \textbf{-} && 0.009 && 0.009 \\
			\hline\hline
			0.8 & 10 && 0.119 && 561.737 && 0.118 && 0.118 \\
			0.8 & 50 && 0.122 && 561.851 && 0.119 && 0.119 \\
			0.8 & 500 && \textbf{-} && \textbf{-} && 0.121 && 0.121 \\
			\hline\hline
			1.6 & 10 && 1.839 && 578.364 && 1.822 && 1.823 \\
			1.6 & 50 && 1.850 && 578.591 && 1.823 && 1.823 \\
			1.6 & 500 && \textbf{-} && \textbf{-} && 1.838 && 1.836 \\
		\end{tabular}
		\caption{Results for $\text{mse}_{\bQ}$.}
		\label{tab_sim_rslp_mseQ}
	\end{table}
	
	\section{Data Examples}\label{sec_data}
	Next we illustrate the algorithm based on two real world data examples, one being the original motivation formulated in the beginning.
	
	\subsection{Orthodont}
	The Orthodont dataset measures the evolution of an orthodontal distance from 27 children over time and additionally contains information about age and gender of the children. Overall the dataset has a total of 108 observations.
	
	We formulate the random intercept model
	\begin{equation*}
	y_{ij} = \beta_0 + \texttt{sex}_i\beta_\text{sex} + \texttt{age}_{ij}\beta_\text{age} + \gamma_{0i} + \varepsilon_{ij}, \quad \gamma_{0i} \sim \mathcal{N}(0, \tau^2)
	\end{equation*}
	where $\texttt{sex}$ is a dummy for female gender and obviously a time-invariant, i.e.\ cluster-constant covariate. The results of \blmm based on 10-fold cross validation and a total $m_\text{stop}$ of 1000 are depicted in Table \ref{tab_orthodont}.
	\begin{table}[h]
		\centering
		\begin{tabular}{ccccccccc}
			&&$\hat{\beta}_0$&&$\hat{\beta}_\text{sex}$&&$\hat{\beta}_\text{age}$ && $\hat{\tau}^2$\\
			\hline
			\texttt{lme} && 17.71 && -2.32 && 0.66 && 3.27 \\
			\blmm && 17.71 && -2.32 && 0.66 && 3.11 \\
			\texttt{bGLMM} && 16.82 && 0.00 && 0.65 && 5.41 \\
		\end{tabular}
		\caption{Estimates for the Orthodont dataset.}
		\label{tab_orthodont}
	\end{table}
	
\begin{figure}[h]
		\centering
		\includegraphics[width = 12cm, trim = 1cm 0 0 0]{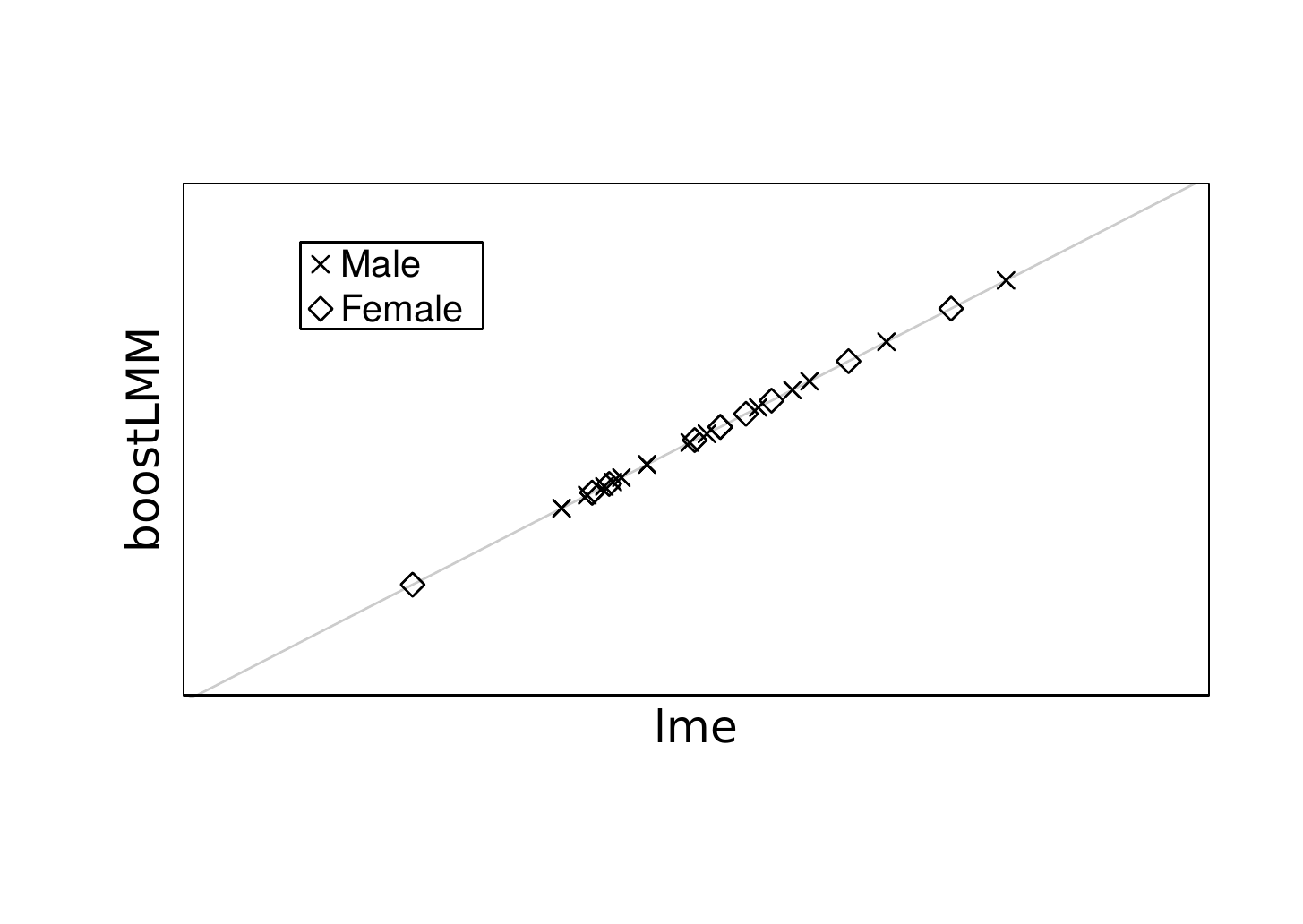}
		\caption{Comparison between random intercept estimates by \texttt{lme} and \texttt{boostLMM}.}
		\label{fig_loesung}
	\end{figure}

	It is evident that \blmm solves the random effects issues occurring with \texttt{bGLMM}. Both the maximum likelihood approach in \texttt{lme} as well as \blmm return matching estimates for fixed and random effects without any shift, which can be seen in Figure \ref{fig_loesung}.
	
	Since the data is very low dimensional, boosting approaches have little to no opportunity to reduce complexity with respect to prediction and usually converge in the regular maximum likelihood estimates without early stopping. Hence, we intend to showcase the method based on another dataset.
	
	\subsection{Primary biliary cirrhosis}
\begin{figure}[h]
		\centering
		\includegraphics[width = 12cm, trim = 1cm 0 0 0]{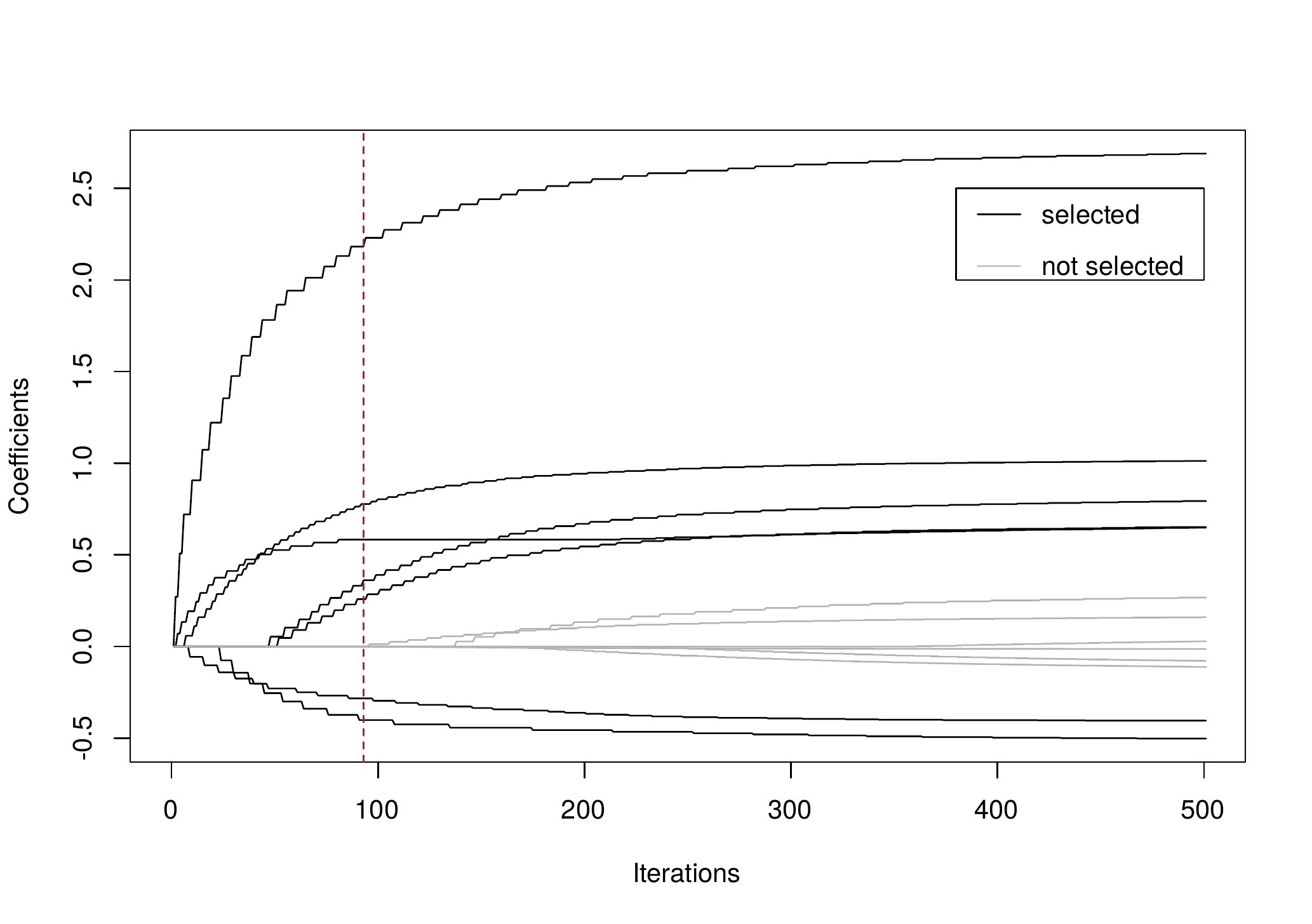}
		\caption{Coefficient paths for model (\ref{eq_pbc_mod}). Based on 10-fold cross validation, the algorithm is stopped after $m_* = 93$ iterations and thus six variables are not included into the model.}
		\label{fig_pbc_coef}
	\end{figure}

	The primary biliary cirrhosis (PBC) data\-set from 1994 \cite{Murtaugh.1994} tracks the change of the serum bilirubin level for a total of 312 PBC patients randomized into a treatment and a placebo group and additionally contains baseline covariates as well as follow-up measurements of several biomarkers. The dataset is, among others, available in the \texttt{JM} package \cite{Rizopoulos.2010} and Table \ref{tab_pbc} gives an overview of the single covariates included in the data and how they are coded in the model formula.
	\begin{table}
		\centering
		\begin{tabular}{cccc}
			\hline
			time-constant & continuous & age at baseline & \texttt{age} \\
			&&&\\
			& discrete & treatment group & \texttt{drug} \\
			&& gender & \texttt{sex} \\
			&& ascites & \texttt{asc} \\
			&& spiders & \texttt{spi} \\
			&& enlarged liver & \texttt{hep} \\
			\hline
			time-varying & continuous & albumin & \texttt{alb} \\
			&& alkaline & \texttt{alk} \\
			&& SGOT & \texttt{SGOT} \\
			&& platelet count & \texttt{pla} \\
			&& prothrombin time & \texttt{pro} \\
			&& time in years & $t$ \\
			\hline
		\end{tabular}
		\caption{Variables of the PBC data set. \texttt{drug} and \texttt{sex} are dummies for treatment group and female gender. Ascites is the abnormal buildup of fluid in the abdomen and spiders are blood vessel malformations in the skin. SGOT is short for serum glutamic oxaloacetic transaminase.}
		\label{tab_pbc}
	\end{table}
	The serum bilirubin level, here modelled as the response variable, is considered a strong indicator for disease progression, hence an appropriate quantification of the impact of the given covariates on the serum bilirubin level will lead to an adequate prediction model for the health status of PBC patients. Using boosting to carry out this quantification will optimize the prediction properties. For $y_{ij}$ denoting the $j$th measurement of serum bilirubin for the $i$th patient, we formulate the random intercept model
	\begin{align}\label{eq_pbc_mod}
	\begin{split}
	y_{ij} &= \beta_0 + \beta_1 \texttt{drug}_i + \beta_2 \texttt{age}_i  + \beta_3 \texttt{sex}_i + \beta_4 \texttt{asc}_i\\
	&\quad + \beta_5 \texttt{hep}_i + \beta_6 \texttt{spi}_i + \beta_7 t_{ij} + \beta_8 t_{ij}^2 + \beta_9 \texttt{alb}_{ij}\\
	&\quad + \beta_{10} \texttt{alk}_{ij} + \beta_{11} \texttt{SGOT}_{ij} + \beta_{12} \texttt{pla}_{ij} + \beta_{13} \texttt{pro}_{ij} + \gamma_{0i} + \varepsilon_{ij}
	\end{split}
	\end{align}
	with $\gamma_{0i} \sim \mathcal{N}(0, \tau^2)$ and an included square time effect, since the effect of time might be nonlinear. Based on 10-fold cross validation, \blmm determined $m_* = 93$ as the best performing number of iterations yielding the corresponding coefficient paths displayed in Figure \ref{fig_pbc_coef}. The coefficient estimates are compared to estimates of a classical \texttt{lme} displayed with the according p-values in Table \ref{tab_pbc_coef}.
	\begin{table}[!t]
		\centering
		\begin{tabular}{cccccc}
			&& \texttt{lme} & p-value && \blmm \\
			\hline
			(Intercept) && 4.15 & 0.00 && 3.88 \\
			\texttt{drug} && -0.48 & 0.26 && -0.40 \\
			\texttt{age} && -0.01 & 0.49 && 0.00 \\
			\texttt{sex} && 0.31 & 0.65 && 0.00 \\
			\texttt{asc} && 2.77 & 0.00 && 2.23 \\
			\texttt{hep} && 0.70 & 0.00 && 0.26 \\
			\texttt{spi} && 0.84 & 0.00 && 0.36 \\
			$t$ && 0.65 & 0.00 && 0.58 \\
			$t^2$ && -0.13 & 0.02 && 0.00 \\
			\texttt{alb} && -0.41 & 0.00 && -0.28 \\
			\texttt{alk} && 0.08 & 0.34 && 0.00 \\
			\texttt{SGOT} && 1.05 & 0.00 && 0.78 \\
			\texttt{pla} && -0.10 & 0.33 && 0.00 \\
			\texttt{pro} && 0.19 & 0.01 && 0.00 \\
			\hline
			$\hat{\tau}$ && 3.55 &&& 3.98
		\end{tabular}
		\caption{Variable selection and shrinkage of \blmm compared to \texttt{lme}.}
		\label{tab_pbc_coef}
	\end{table}
	
	The variables age, gender, alkaline, platelets, prothrombin as well as the squared time were not selected and thus $\hat{\beta}_k = 0$ for $k = 2,3,8,10,12,13$. Please note that those variables are also not significant in the \texttt{lme} model. The boosting model, however, has the advantage of leading to better prediction. This is ensured by not entering the variables, which do not have any explanatory power into the model at all. A further advantage is the shrinkage of the variables: while being similar to the values of the \texttt{lme} model, the parameters in the boosting model do not lead to overfitting.

	\section*{Discussion}
	The updated algorithm is due to its minor and major tweaks capable of dealing with cluster-constant covariates in linear mixed models by preventing the random effects from taking up too much space. In addition, it preserves the well-known advantages of boosting techniques in general by offering variable selection and a good functionality even in high dimensional setups. As a very important side effect the computational effort receives a tremendous decrease making the algorithm more applicable to real world scenarios.
	
	The slightly underperforming variable selection properties compared to \texttt{GMMBoost} are due to the current stopping iteration being determined by cross validation, which does not originally address variable selection. Alternatives include relying on information criteria like AIC or BIC, where it is sufficient to formulate a global hat matrix incorporating the proposed changes of the algorithm, or established tools for variable selection in boosting procedures like probing \cite{Hepp.2017} or stability selection \cite{Meinshausen.2010}.
	
	Canonical extensions of the successful concept include incorporating non-linear predictor functions, i.e.\ estimation of smooth effects based on P-splines or extending the algorithm from linear mixed models to generalized mixed models to allow more flexible inference for a wider class of data structures. Both have been incorporated in \cite{Groll.Diss} for classical likelihood-based boosting and it is assumed that the proposed tweaks in the present work would improve performance of the more flexible approaches as well.
	
	\bibliographystyle{amsplain}
	\bibliography{bibliothekus.maximus}
	
\end{document}